\documentclass[prx,aps,twocolumn]{revtex4-1}
\usepackage[utf8]{inputenc}
\usepackage{graphics, bm, psfrag, amsmath, amssymb, epsfig, grffile, float, bbold}
\usepackage{dsfont}
\usepackage{color}
\usepackage{hyperref}
\usepackage{amsfonts,amsthm,mathtools}
\usepackage{latexsym,graphicx}
\usepackage{amscd}
\usepackage{natbib} 
\usepackage{xcolor}

\usepackage[utf8]{inputenc}

\newcommand{\ee}{{\rm e}}

\newcommand{\ii}{\mathrm{i}}

\newcommand{\om}{\omega} 
\newcommand{\re}{\mathrm{Re}}

\newcommand{\rj}{\mathrm{j}}
\newcommand{\cV}{\mathcal{V}}

\newcommand{\eps}{\epsilon}

\newcommand{\R}{{\mathbb R}}

\newcommand{\Z}{{\mathbb Z}}

\newcommand{\cF}{\mathcal{F}}

\begin{document}

\title{Closed-form propagator of the Calogero model}

\author{Valdemar Melin}
\author{Edwin Langmann}

\affiliation{Department of Physics, KTH Royal Institute of Technology, SE-106 91 Stockholm, Sweden}

\affiliation{Nordita, KTH Royal Institute of Technology and Stockholm University, SE-106 91 Stockholm, Sweden}

\date{March 12, 2023} 

\begin{abstract} 
We present an exact closed-form expression for the propagator of the Calogero model, i.e., for the integral kernel of the time evolution operator of the quantum many-body system on the real line with an external harmonic potential and inverse-square two-body interactions. 
This expression is obtained by combining two results: first, a simple formula relating this propagator to the eigenfunctions of the Calogero model without harmonic potential and second, a formula for these  eigenfunctions as finite sums of products of polynomial two-body functions. 
\end{abstract}
\maketitle

\noindent {\bf Introduction.} 
Integrable systems have always played an important role in physics, with historic examples including the Kepler problem and the quantum mechanical model of hydrogen.
Today, quantum many-body systems of Calogero-Moser-Sutherland (CMS) type \cite{calogero1971,moser1976,sutherland1972,toda1967,olshanetsky1976,olshanetsky1983}  
have become prime  integrable models due to their  successful applications in many different  areas of physics,  including (quantum) hydrodynamics, gauge theories, solitons, fractional quantum Hall effect, conformal field theory, and non-equilibrium physics; see e.g.\    \cite{polychronakos1995,dhoker1998,abanov2005,bettelheim2006,agarwal2006,stone2008,wiegmann2012,estienne2012,isachenkov2016,berntson2020,spohn2023,berntson2023}. 
Furthermore, the intriguing mathematics behind  these models has received attention by mathematical physicists and mathematicians with specializations as diverse as representation theory, differential geometry, combinators, and the modern theory of special functions; see e.g.\ \cite{ruijsenaars1999,polychronakos2006,chalykh2008,etingof2007,nekrasov2010,vandiejen2012}. 
However, despite of all this work, we are still far from knowing {\em all} quantities of interest in physics for CMS-type models analytically. 
One important such example is the propagator, i.e., the integral kernel of the time evolution operator in position representation. 

Numerical solutions of the time-dependent Schr\"odinger equation are typically restricted to small particle numbers and, for this reason, exact propagators are of interest to the large community of physicists trying to understand the dynamics of real quantum systems. 
Unfortunately, as for today, few examples of exact propagators are known in closed form; these examples include the heat kernel for the non-interacting case, the Mehler kernel for the quantum harmonic oscillators, and  and the Hille-Hardy formula for the propagator of the singular harmonic oscillator; see e.g.\ \cite{khandekar1975,nowak2013,sakurai2014}. 
In this paper, we present exact analytic formulas for the propagator of the quantum many-body system studied by Calogero in his seminal paper \cite{calogero1971}; see \eqref{explicit}--\eqref{C3} for one way to write our result in closed form. 
Since the Calogero model is a prototype for interacting systems in one dimension, we believe that our results are of general interest in physics. 
Furthermore, our exact results for the Calogero model suggest a new kind of ansatz to find approximate wave functions for generic quantum many-body systems taking into account non-trivial correlations (this is discussed in our conclusions). 
We also  hope that our results can inspire future work finding exact propagators for other CMS-type quantum systems. 

\medskip 

\noindent {\bf Calogero model.} 
The Calogero model is a many-body generalization of the quantum harmonic oscillator to an arbitrary number, $N$, of particles with inverse-square two-body interactions; it  is defined by the quantum mechanical Hamiltonian ($i<j$ is short for $1\leq i<j\leq N$) 
\begin{equation} 
\label{H}
H = \sum_{i=1}^{N}  \frac12\left( -\frac{\partial^2}{\partial x_i^2} + \om^2x_i^2 \right) + \sum_{i<j} \frac{\ell(\ell+1)}{(x_i-x_j)^2}
\end{equation} 
where $x_i\in\R$ are the particle positions, and $\om\geq 0$ and $\ell\geq -1/2$ are parameters determining the strength of the harmonic potential and two-body interactions, respectively (we use units such that $m=\hbar=1$).

We recall that the eigenfunctions of  $H$ of interest in physics have the form
\begin{equation} 
\label{eigenfunctions} 
\Psi(x) =\ee^{-\om x^2/2}\cV_N(x)^{\ell+1}P(x) 
\end{equation} 
with $P(x)$ 
symmetric and analytic functions in the variables $x=(x_1,\ldots,x_N)\in\R^N$, $x^2$ short for $\sum_{i=1}^Nx_i^2$,  and  $\cV_N(x)\coloneqq \prod_{i<j}(x_i-x_j)$ the Vandermonde. 
In particular, since the Vandermonde changes sign under particle exchanges $x_1\leftrightarrow x_2$ etc., the Calogero models describes fermions and bosons for even and odd integers $\ell$, respectively; for this reason, we restrict ourselves to integer $\ell\geq 0$.
For $\om>0$, the  functions $P(x)$ are symmetric polynomials which are natural many-variable generalizations of the Hermite polynomials \cite{lassalle1991,brink1992,baker1997,nekrasov1997,rosler1998,feigin2021}; in this case, the Calogero model has discrete energy eigenvalues. 
The case $\om=0$ is special in that the Calogero Hamiltonian has continuous spectrum and the eigenfunctions are generalizations of (anti-)symmetrized plane waves; as will be seen, this case is important for us. 
 
The propagator of the Calogero model is the time evolution operator in position representation for $H$ in \eqref{H}, $K_N(x,y;t)\coloneqq \langle x|\ee^{-\ii Ht}|y\rangle$, using the Dirac bra-ket notation. 
Mathematically, this propagator can be defined as integral kernel in the following solution of the time dependent Schr\"odinger equation 
$\ii\partial_t\psi(x;t)=H\psi(x;t)$, 
\begin{equation} 
\label{psi} 
\psi(x;t) = \int_{\R^N} K_N(x,y;t)\psi_0(y) d^N y
\end{equation} 
with $\psi_0(x)\coloneqq \psi(x;0)$ an arbitrary wave function at initial time $t=0$ \cite{remark2}. 

Our explicit formula for this propagator in \eqref{explicit} below is obtained by combining two results: 
(i) A relation between $K_N(x,y;t)$ and the eigenfunctions $\Psi_N(x;p)$ of the Calogero Hamiltonian $H|_{\om=0}$ in \eqref{H} for $\om=0$ (Proposition 1), (ii) a simple explicit formula for $\Psi_N(x;p)$ (see Proposition~2 for $N=3$ and \eqref{PsiN} with \eqref{cFN} in general). 

\medskip 

\noindent {\bf Relation to wave functions.} To motivate our first result, we recall the well-known propagator for the quantum harmonic oscillator $H|_{N=1}$ known as Mehler kernel, 
\begin{equation} 
\label{Mehler} 
K_1(x,y;t) = \frac{\ee^{\ii\om[(x^2+y^2)\cos(\om t)-2xy]/2\sin(\om t)}}{(2\pi\ii \sin(\om t)/\om)^{1/2}}
\end{equation} 
($x,y\in\R$); see e.g.\ \cite[Eq.\ (2.6.18)]{sakurai2014}. From this, one can obtain the propagator of the Calogero model in the non-interacting case $\ell=0$, which describes non-interacting fermions in a harmonic potential, as follows: multiply the product of $N$ Mehler kernels for the variables $x_i$, $i=1,\ldots,N$, and anti-symmetrize, i.e., 
\begin{equation}
\label{MehlerN} 
K_N(x,y;t)|_{\ell=0}= \frac1{N!}
\sum_{\sigma\in S_N} \eps_\sigma\prod_{i=1}^N K_1(x_{\sigma(i)},y_i;t)
\end{equation} 
with $S_N$ is the permutation group and $ \eps_\sigma=+1$ and $-1$ for even and odd permutations $\sigma\in S_N$, respectively.  
We note that \eqref{MehlerN} can be written as 
\begin{equation} 
\label{K} 
K_{N}(x,y;t) = \frac{\ee^{\ii \om(x^2+y^2)/2\tan(\om t)}}{(2\pi\ii \sin(\om t)/\om)^{N/2}}
\Psi_{N}\left(x;\frac{-\om y}{\sin(\om t)} \right) 
\end{equation} 
with $\Psi_{N}(x;p)|_{\ell=0} = (1/N!)\sum_{\sigma \in S_N} \eps_\sigma \ee^{\ii p\cdot x_{\sigma}}$ the plane-wave 
 eigenfunction of the Hamiltonian $H_{\ell=0,\om=0}$ describing non-interacting fermions on the real line without harmonic potential (we use the short-hand notation $x_\sigma$ and $p\cdot x$ for $(x_{\sigma(1)},\ldots,x_{\sigma(N)})$ and $\sum_{i=1}^N p_ix_i$, respectively).
We found that the propagator of the Calogero model is given by \eqref{K} even for non-zero $\ell$ but where $\Psi_N(x;p)$ is the eigenstate of the corresponding Calogero Hamiltonian $H|_{\om=0}$  without harmonic potential. 
To be precise (see the Appendix for proof): 

\medskip 

\noindent {\bf Proposition~1:}  {\em The propagator of the Calogero model is given by \eqref{K} with the wave function $\Psi_{N}(x;p)$ satisfying the following conditions:  
(i) it solves the stationary Schr\"odinger equation 
\begin{equation} 
\label{HPsi=EPsi}
H|_{\om=0}\Psi_{N}(x;p)=(p^2/2)\Psi_{N}(x;p),
\end{equation} 
(ii) it converges to $(1/N!)\sum_{\sigma \in S_N} \eps_\sigma^{\ell+1}\ee^{\ii p\cdot x_{\sigma}}$ in the limit $|x_{i}-x_{i+1}|\to \infty$, $i=1,\ldots,N-1$, where all particles are infinitely far apart, 
(iii) $\Psi_{N}(x;p)/\cV_N(x)^{\ell+1}$ is analytic and symmetric in the variables $x$, (iv) $\Psi_{N}(x;p)$ satisfies the scaling relations
\begin{equation} 
\label{s} 
\Psi_{N}(sx;p) =\Psi_{N}(x;sp)\quad (s>0). 
\end{equation} 
} 

\medskip 

We believe this result deserves to be generally known: the wave function $\Psi_N(x;p)$, $p\in\R^N$,  appearing in the exact formula \eqref{K} for the propagator of the Calogero model is identical with the eigenfunctions of the translationally invariant Calogero Hamiltonian $H|_{\om=0}$  of interest in physics.
It is known that the conditions (i)--(iii) above determine these eigenfunctions uniquely; as shown in Appendix~A, (iv) is implied by (i)--(iii). 

\medskip 

\noindent {\bf Explicit wave functions.} 
There exist several explicit formulas for the wave function $\Psi_N(x;p)$ in the literature (such results can be found in \cite{calogero1969,chalykh1990,opdam1993,chalykh1999,felder2009}, for example); any of these provides an  explicit formula for the propagator $K_N(x,y;t)$ by \eqref{K}. 
We found a formula for this wave function which has a natural physics interpretation and is particularly simple from a computational point of view: 
\begin{equation}
\label{PsiN} 
\Psi_N(x;p) = \frac1{N!}\sum_{\sigma\in S_N} \eps_\sigma^{\ell+1} \cF_N(x_\sigma;p) \ee^{\ii p\cdot x_\sigma}
\end{equation} 
with $\cF_N(x;p)$ a finite(!) linear combination of products of the polynomials 
\begin{equation}
\label{Fk}  
F_k(X)\coloneqq \sum_{a=k}^\ell \frac{(\ell+a)!}{(\ell-a)!(a-k)!}\frac1{X^{a}} \quad (k=0,\ldots,\ell) 
\end{equation} 
in the variables $1/X=1/X_{i,j}$ with 
\begin{equation} 
\label{Xjk} 
X_{i,j}\coloneqq -\ii (p_i-p_j)(x_i-x_j)\quad (1\leq i<j\leq N) ;
\end{equation} 
see \eqref{cFN} below for general $N$ and \eqref{cF3} for $N=3$. 

To motivate this result and explain the significance of the polynomials in \eqref{Fk}, we recall that the known eigenfunctions in the two-body case $N=2$ are given by  \cite{calogero1969}
\begin{equation} 
\label{Psi2} 
\Psi_2(x;p) = 
\ee^{\ii(x_1+x_2)(p_1+p_2)/2}(-\ii)^{\ell+1}z\rj_{\ell}(z)
\end{equation} 
with $z\coloneqq (p_1-p_2)(x_1-x_2)/2$, where $\rj_\ell(z)$ are spherical Bessel functions \cite{DLMF}. 
One can verify that \eqref{Psi2} can be written as in \eqref{PsiN} with  
\begin{equation} 
\label{cF2} 
\cF_2(x;p)=F_0(X_{1,2})
\end{equation} 
the function in \eqref{Fk} for $k=0$ and with $X_{1,2}$ in \eqref{Xjk} (using e.g.\ \cite[Eq.\ 10.49.2]{DLMF}) \cite{remark3}. 
Thus, $F_0(X)$ accounts for the two-body correlations in the Calogero model for $\om=0$. Since $F_0(X)\to 1$ as $X\to\infty$, $\Psi_2(x;p)$ simplifies to symmetrized or anti-symmetrized plane waves in the limit $|x_1-x_2|\to\infty$ when the particles are infinitely apart, as expected. 
The other functions in \eqref{Fk} are descendants of $F_0(X)$ in the following sense, $F_k(X)=\partial_s^{k}F_0(X/s)|_{s=1}$ for $k=1,\ldots,\ell$. 
Moreover, the larger $k$, the faster $F_k(X)$ decays as $X\to\infty$: $F_k(X)=(\ell+k)!/(\ell-k)!X^k+\ldots$ with the dots indicating terms vanishing faster as $X\to \infty$.  
For example, for $\ell=1$ we have $F_0(X)=1+2/X$ and $F_1(X)=2/X$, and for $\ell=2$ these functions are $F_0(X)=1+6/X+12/X^2$, $F_1(X)=6/X+24/X^2$, and $F_2(X)=24/X^2$, etc. 

In the $N$-body case and in the asymptotic region when all particles are infinitely far apart, the two-body interactions are irrelevant and the eigenfunction $\Psi_N(x;p)$ can be obtained by symmetrizing or anti-symmetrizing a product of $N$ one-particle eigenfunctions; this corresponds to \eqref{PsiN} with $\cF_N(x;p)\to 1$ in the asymptotic region. 
More generally, the function $\cF_N(x;p)$ must satisfy {\em cluster properties} as a consequence of the fact that interactions vanish for particles infinitely far apart: if we take the limit where we divide the particles into $K$ groups of $N_J$ particles, $J=1,\ldots,K$ and $\sum_{J=1}^K N_J=N$,  and such that  particles in the same group keep finite distances but particles in different groups become infinitely apart, the interactions between particles in different groups become irrelevant and   
 \begin{equation} 
 \label{clusterproperty} 
 \cF_N(x;p) \to \prod_{J=1}^K \cF_{N_J}(x_J;p_J) 
 \end{equation} 
 where $x_J\in\R^{N_J}$ and $p_J\in\R^{N_J}$ are the positions and momenta of the particles in the group $J$ ($\cF_1= 1$). 
 Clearly, \eqref{clusterproperty} is a highly restrictive constraint on the function $\cF_N(x;p)$. For example, by choosing all $N_J$ to be $2$ or $1$ one finds that  \eqref{clusterproperty} implies 
 \begin{equation}
\label{cFNapprox} 
\cF_N(x;p) = \prod_{1\leq i<j\leq N} \cF_2((x_i,x_j);(p_i,p_j))  + \cdots 
\end{equation} 
where the dots indicate sub-leading terms. 

The considerations in the previous paragraph are general and apply to {\em any} translationally invariant quantum-many body systems with two-body interactions that vanish in the limit when particles distances become infinite. It thus is not surprising that \eqref{cFNapprox} holds true for the Calogero model with $ \cF_2((x_i,x_j);(p_i,p_j))=F_0(X_{i,j})$. 
What is special for the Calogero model is that the sub-leading terms in \eqref{cFNapprox} can be also written as linear combinations of products of two-body functions in \eqref{Fk}: we found that the function $\cF_N(x;p)$ in \eqref{PsiN} has the exact form  
\begin{equation} 
\label{cFN} 
\cF_N(x;p) = \sum_{\mathbf{k}} c_N({\mathbf{k}}) \prod_{1\leq i<j\leq N} F_{k_{i,j}}(X_{i,j}) 
\end{equation} 
with coefficients $c_N(\mathbf{k})$ depending on the integer vectors $\mathbf{k}=(k_{i,j})_{1\leq i<j\leq N}$ in the finite set $\{0,\ldots,\ell\}^{N(N-1)/2}$. 
We believe there is a simple general formula for the coefficients $c_N(\mathbf{k})$; at this point, we know this formula in some special cases, including $N=3$ and arbitrary integers $\ell>0$: 

\medskip 

\noindent {\bf Proposition 2.} {\em For $N=3$, the eigenfunctions of $H|_{\om=0}$ characterized in Proposition~1 are given by \eqref{PsiN} and 
\begin{equation} 
\label{cF3} 
\cF_3(x;p) = \sum_{k=0}^\ell \frac{(\ell-k)!}{(\ell+k)!k!}F_k(X_{1,2})F_k(X_{1,3})F_k(X_{2,3})  
\end{equation} 
with $F_k(X)$ in \eqref{Fk} and $X_{i,j}$ in \eqref{Xjk}.
}
\medskip

We discovered this formula by computer experiments and subsequently found a mathematical proof for all $\ell$; since this proof is technical and it is easy to  verify the result for $\ell=1,\ldots,10$ using symbolic computing software on a laptop, we plan to publish this proof elsewhere. 
We also obtained explicit results for larger values of $N$ for the simplest non-trivial case $\ell=1$; see Appendix~B(ii). 

\begin{figure}
\hbox{\hspace{-2.3em} \includegraphics[width=0.6\textwidth]{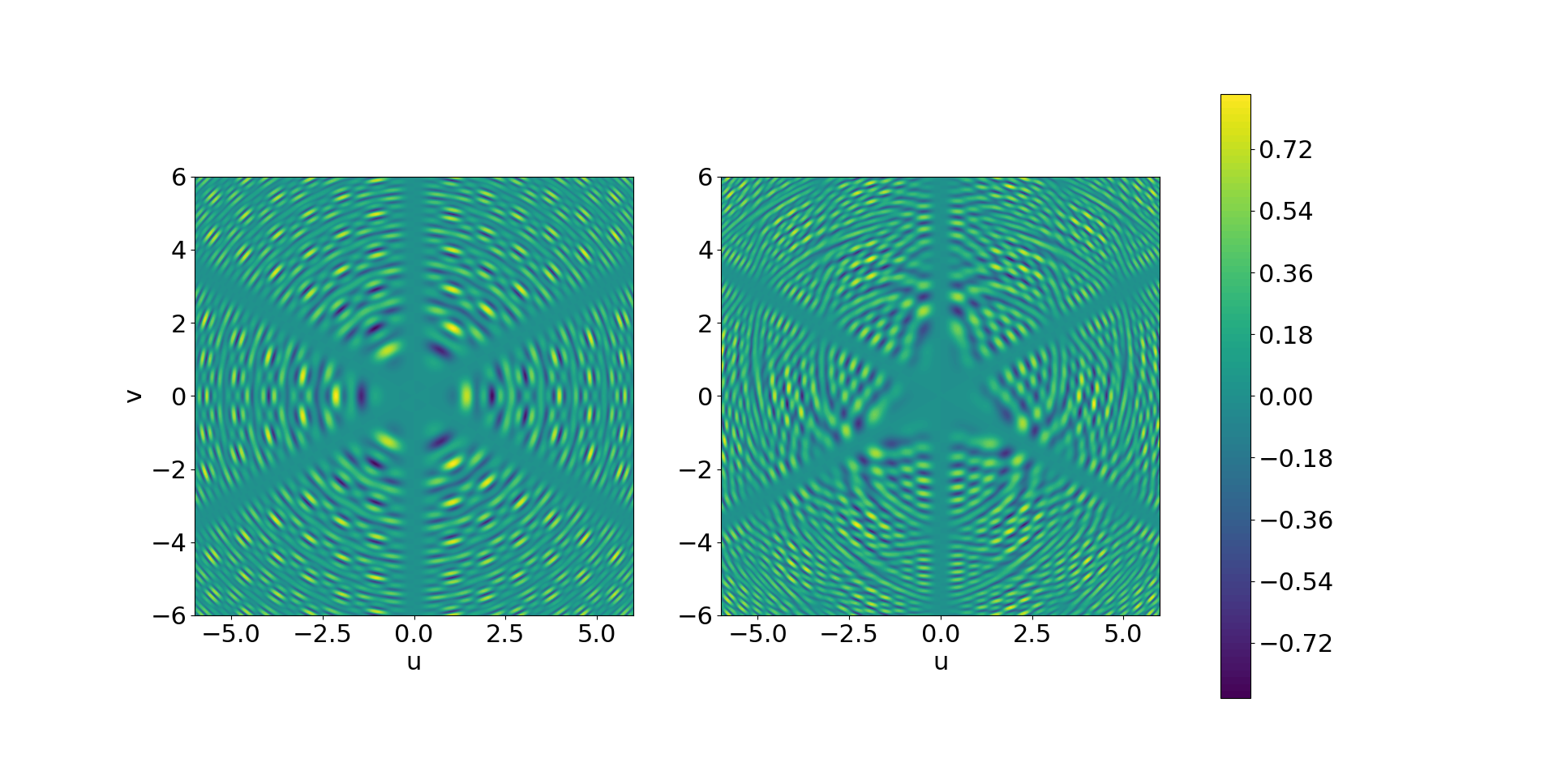} }
\caption{Propagator of the Calogero model for $(N,\ell)=(3,2)$: real part of $K_3(x,y;t)$ for $x_1+x_2+x_3=0$ as a function of $u=2^{-\frac12}(x_1-x_2)$, $v=6^{-\frac12}(x_1+x_2-2x_3)$  \cite{calogero1969} at time $t=\pi/16$ in units $\omega=1$; LHS: $y=(-1,0,1)$, RHS: $y=(-1,-0.5,1.5)$.}
\label{Fig1}
\end{figure}

Clearly, by combining Propositions~1 and 2, we obtain a fully explicit propagator for the Calogero model for the case $N=3$. Moreover, by finding an explicit formula for the coefficients $c_N(\mathbf{k})$ in \eqref{cFN}, this can be extended to other particle numbers $N$. 
By combining cluster property \eqref{clusterproperty} with the results \eqref{cF2} and \eqref{cF3} for $N=2$ and $N=3$, respectively, many (but not all) of the coefficients $c_N(\mathbf{k})$ can be determined. 
Fig.~\ref{Fig1} shows two examples of visualizations of the propagator for $(N,\ell)=(3,2)$. 

\medskip 

\noindent {\bf Conclusions.} We present exact analytic results for the Calogero model providing the following  many-body generalization of the Mehler kernel $K_1$ in \eqref{Mehler}, 
\begin{multline} 
\label{explicit} 
K_N(x,y;t) = 
\frac1{N!}\sum_{\sigma\in S_N} \eps_\sigma^{\ell+1} \prod_{i=1}^N K_1(x_{\sigma(i)},y_i;t) \\ \times 
\sum_{\mathbf{m}} C_{N}(\mathbf{m}) \prod_{ i<j}\Bigg( \frac{-\ii\sin(\omega t)/\omega}{(x_{\sigma(i)}-x_{\sigma(j)})(y_i-y_j)}\Bigg)^{m_{ij}}
\end{multline} 
with $\mathbf{m}=(m_{i,j})_{1\leq i<j\leq N}\in \{0,\ldots,\ell\}^{N(N-1)/2}$ and postive-integer coefficients $C_N(\mathbf{m})$ which,  for $N=2$ and $N=3$, are given by
($(a,b,c)$ is short for $(m_{12},m_{13},m_{23})$), 
\begin{equation} 
\label{C2} 
C_{2}(a) = \frac{(\ell+a)!}{(\ell-a)!a!}
\end{equation} 
and 
\begin{multline} 
\label{C3} 
C_{3}(a,b,c) = C_2(a)C_2(b)C_2(c) \\ \times 
\sum_{k=0}^{\min(a,b,c)} \frac{(\ell-k)!a!b!c!}{(\ell+k)!k!(a-k)!(b-k)!(c-k)!}, 
\end{multline} 
respectively. We expect that similar positive integer formulas for $C_N(\mathbf{m})$ exist for all $N=4,5,\ldots$; see Appendix~B(ii) for results on  $C_N(\mathbf{m})$ for $\ell=1$ and arbitrary $N$. 

Our closed-form expressions for the propagator of the Calogero model provide analytic tools to compute the dynamics of a prototype quantum many-body system with non-trivial interactions.
We hope that this will find applications in, for example, non-equilibrium physics or realizations of the Calogero model in cold atom systems \cite{polkovnikov2011}.

One important ingredient to our result is a formula for the eigenfunctions of the Calogero model without harmonic potential which is of interest in its own right; see \eqref{PsiN} and Proposition~2.
In particular, we  propose to use this exact result as a guide to find approximate wave functions for other quantum many-body systems.
The simplest example would be \eqref{cFNapprox} with $\cF_2(x;p)$ obtained by solving the two-body problem (ignoring the dots) or, alternatively, one could try to find a mean-field type equation determining $\cF_2(x;p)$. 
More generally, one could use \eqref{cF3} as an ansatz for the three-body wave function with $F_k(X_{i,j})$ replaced by variational two-body functions $\cF_2^{(k)}((x_i,x_j);(p_i,p_j))$, $k=0,\ldots,\ell$,  and then use \eqref{cFN} for the many-body wave functions: in this way, one could try to systematically improve the approximate wave functions by increasing $\ell$ and having the  cluster properties \eqref{clusterproperty} always satisfied. 

Our result for $N=2$ was known before: in this case, the propagator given in \eqref{K} and \eqref{Psi2} is a product of a Mehler kernel for the center-of-mass $x_1+x_2$,  and the propagator for the singular harmonic oscillator given by the Hille-Hardy formula \cite{nowak2013} for the relative coordinate $x_1-x_2$. 
We also mention an integral representation of the propagator of the Calogero model in the special case $\ell=-1/2$ obtained in Ref.~\cite{fleury1998}. 
Proposition~2 has relations to previous results in the mathematics literature \cite{chalykh1990,opdam1993,chalykh1999,felder2009,noumi2012,kazhdan1978,hallnas2015}, as discussed in more detail in Appendix~C(v) and (vi). 
Clearly, it would be interesting to extend Proposition~2 to all particle numbers $N>3$.

The interested reader can find further results and further remarks on mathematical aspects of our work in Appendix~B and C, respectively.

\smallskip

\noindent {\bf Acknowledgements.}  
We would like to thank Martin Halln\"as and Masatoshi Noumi for helpful discussions.
We are grateful to Martin Halln\"as for pointing out Ref.~\cite{rosler1998} containing results coming close to our Proposition~1. 
E.L. acknowledges support from the European Research Council, Grant Agreement No.\ 2020-810451.

\medskip

\noindent{\bf Appendix A: Proof of Proposition~1.} We first show that properties (i)--(iii) of the eigenfunction $\Psi_N(x;p)$ imply property (iv): Replacing $x\to sx$ ($s>0$) changes $H|_{\om=0}\to s^{-2}H|_{\om=0}$. Therefore, (i) implies $s^{-2}H|_{\om=0}\Psi_N(sx;p)=(p^2/2)\Psi_N(sx;p)$; clearly, this equation is also satisfied by $\Psi_N(x;sp)$. 
Moreover, (ii) and (iii) are satisfied by $\Psi(sx;p)$ if and only if they are satisfied by $\Psi(x;sp)$. Since (i)--(iii) have a unique solution, (iv) follows. 
We note in passing that (iii) is a technical condition ruling out unphysical solutions.

To verify that $\psi$ in \eqref{psi} satisfies the pertinent time-dependent Schr\"odinger equation, we write $K_N=G\Psi_N$ with $G$ short for $\ee^{\ii\om(x^2+y^2)/2\tan(\om t)}/(2\pi\ii \sin(\om t)/\om)^{N/2}$ and $\Psi_N=\Psi_N(x;p)$ with $p=-\om y/\sin(\om t)$, suppressing arguments. 
We compute, using the product and chain rules of differentiation, 
\begin{multline} 
\label{computation}
(\ii\partial_t - H) K_N = G\left( (p^2/2)- H|_{\om=0}\right)\Psi_N \\
+ G \frac{\ii\om}{\tanh(\om t)}\sum_{i=1}^N\left(x_j\frac{\partial\Psi_N}{\partial x_j} - p_j\frac{\partial\Psi_N}{\partial p_j}  \right)\\
+  \Psi_N\left( \ii\partial_t -H|_{\ell=0} -(p^2/2)\right)G; 
\end{multline} 
we added and subtracted the same term $(p^2/2)K_N$ in the first and last lines, respectively; 
the first and second term in the second line come from $\nabla G\cdot \nabla \Psi_N$ and $G\ii\partial_t\Psi_N$, respectively, using $\ii\partial_t p_j = -\ii\om p_j/\tan(\om t)$;  
we included the two-body and harmonic potential terms in the first and third lines, respectively. 
By differentiating \eqref{s} with respect to $s$ and setting $s=0$ we obtain an identity showing that the terms in the second line in \eqref{computation} vanish.
We now use the fact that, in the non-interacting case $\ell=0$, Proposition~1 is true by \eqref{Mehler} (this is explained in the main text). 
This implies that $(\ii\partial_t -H|_{\ell=0})G\Psi_N|_{\ell=0}=0$. 
Thus, setting $\ell=0$ in \eqref{computation} and using that $H|_{\om=\ell=0}\Psi_N|_{\ell=0}=(p^2/2)\Psi_N|_{\ell=0}$, we find that $\left( \ii\partial_t -H|_{\ell=0} -(p^2/2)\right)G=0$; 
this proves that the term in the third line in \eqref{computation} vanishes as well.  
Thus, if \eqref{HPsi=EPsi} and \eqref{s} hold true, then $(\ii\partial_t - H) K_N=0$, and this implies that $\psi$ in \eqref{psi} is a solution of the time dependent Schr\"odinger equation $\ii\partial_t\psi=H\psi$.

In the limit $t\to 0$ of \eqref{K}, one can replace the functions $\sin(\om t)/\om$ and $\tan(\om t)/\om$ by $t$ and $\Psi_N(x;-\om p/\sin(\om t))$ by $(1/N!)\sum_{\sigma \in S_N} \eps_\sigma^{\ell+1}\ee^{-\ii \cdot x_{\sigma}\cdot y/t}$, respectively; the latter follows from (ii). 
Thus, to compute the limit $t\to 0$ of $\psi$ in \eqref{psi}, we can replace $K_N(x,y;t)$ in \eqref{K} by  
\begin{multline}
\label{freepropagator}
\frac{\ee^{\ii(x^2+y^2)/2t}}{(2\pi\ii t)^{N/2}}\frac1{N!}\sum_{\sigma \in S_N}  \eps_\sigma^{\ell+1}\ee^{-\ii \cdot x_{\sigma}\cdot y/t}\\= 
\frac1{N!}\sum_{\sigma \in S_N}  \eps_\sigma^{\ell+1} \frac{\ee^{\ii(x_\sigma -y)^2/2t}}{(2\pi\ii t)^{N/2}}; 
\end{multline} 
this is the well-known propagator of free fermions/bosons for even/odd $\ell$, respectively, converging to the appropriate Dirac delta in the limit $t\to 0$. 
Thus, $\psi$ in \eqref{psi} satisfies $\psi(x;t)\to \psi_0(x)$ as $t\to 0$.  

\medskip

\noindent{\bf Appendix B: Further results.} We give a further visualizations of our result in (i), and in (ii) we present results on the coefficients $C_N(\mathbf{m})$ in \eqref{explicit} for $\ell=1$ and $N=4,5,\ldots$. 

\medskip

\begin{figure}
\hbox{\hspace{-2.3em} \includegraphics[width=0.6\textwidth]{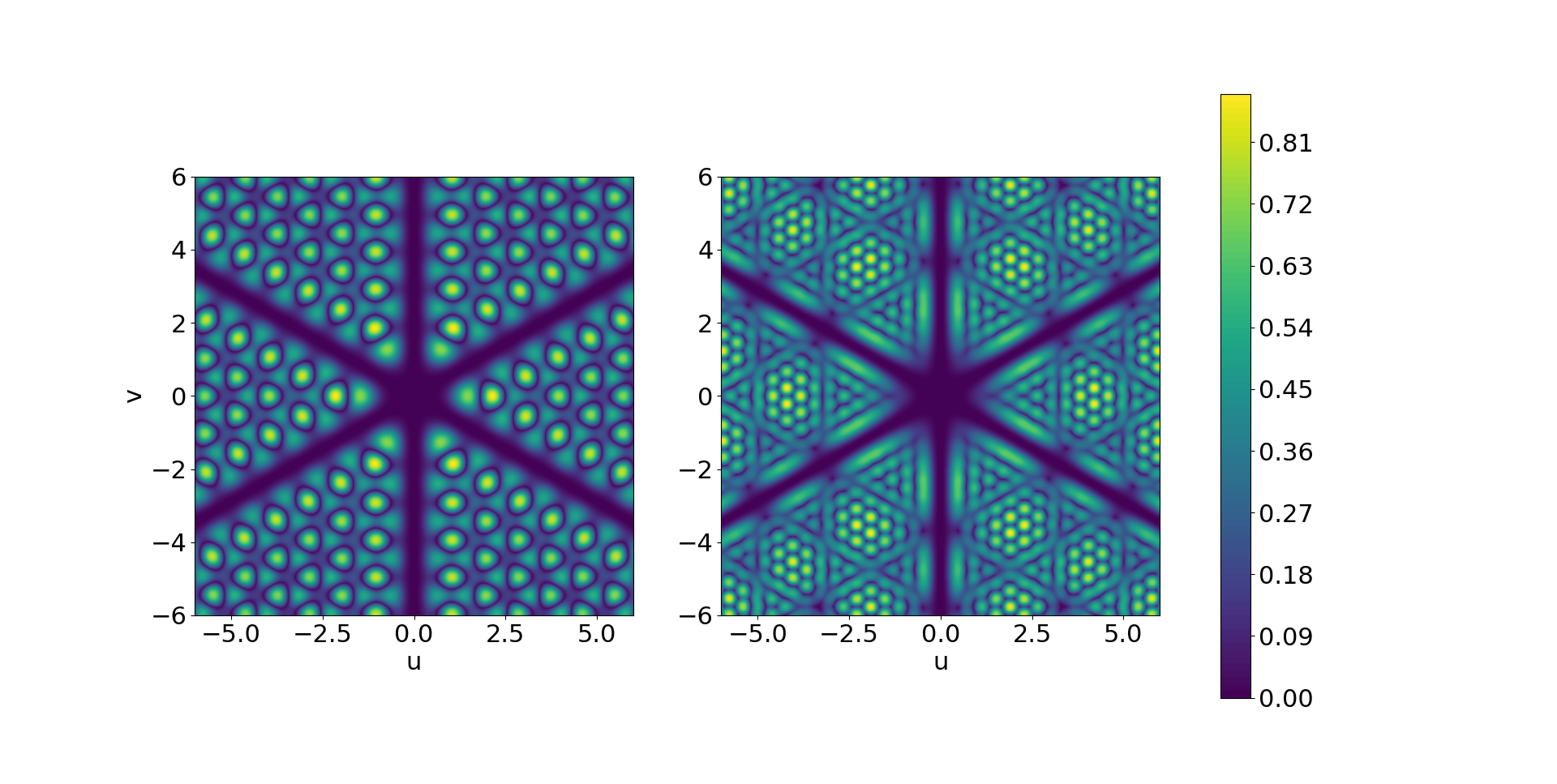} }
\caption{Propagator of the Calogero model for $(N,\ell)=(3,2)$: Absolute value of $K_3(x,y;t)$ for $x_1+x_2+x_3=0$ as a function of $u=2^{-\frac12}(x_1-x_2)$, $v=6^{-\frac12}(x_1+x_2-2x_3)$  at time $t=\pi/16$ in units $\omega=1$; LHS: $y=(-1,0,1)$, RHS: $y=(-1,-0.5,1.5)$.}
\label{Fig2}
\end{figure}

\noindent {\bf (i)} Fig.~\ref{Fig2} shows a visualization of the propagator for $(N,\ell)=(3,2)$ which is complementary to the one in Fig.~\ref{Fig1}: 
while Fig.~\ref{Fig1} shows $\re K_3(x,y;t)$ for particular values of $y=(y_1,y_2,y_3)$ and $t$ as a function of the relative coordinates $u=2^{-\frac12}(x_1-x_2)$, $v=6^{-\frac12}(x_1+x_2-2x_3)$  for fixed center-of-mass $x_1+x_2+x_3=0$ (we follows Calogero \cite{calogero1969} in our choice of coordinates), Fig.~\ref{Fig2}  shows $|K_3(x,y;t)|$ for these two examples (with everything else the same). 
To set this result in perspective we note that, for $N=1$ (Mehler kernel), $|K_1(x,y;t)|$ is a time dependent constant (independent of $x$ and $y$), i.e.,\ the non-trivial structure in Fig.~\ref{Fig2} is due to many-body effects. 
We note in passing that the dark regions in Fig.~\ref{Fig2} containing the straight lines $u=0$ and $v=\pm u/\sqrt{3}$ are a consequence of $|K_3(x,y;t)|$ vanishing like $|\cV_3(x)|^3$ as $x_i\to x_j$ ($1\leq i<j\leq 3$). 

\medskip

\noindent {\bf (ii)} For $\ell=1$, we computed the coefficients $C_N(\mathbf{m})$ for all $N=2,\ldots,5$ using symbolic computer software.  
As explained below, our results suggest that the coefficients $C_N(\mathbf{m})$ have a combinatorial interpretation. 

\begin{figure}
\hbox{\hspace{0em}\vspace{0em} \includegraphics[width=0.45\textwidth]{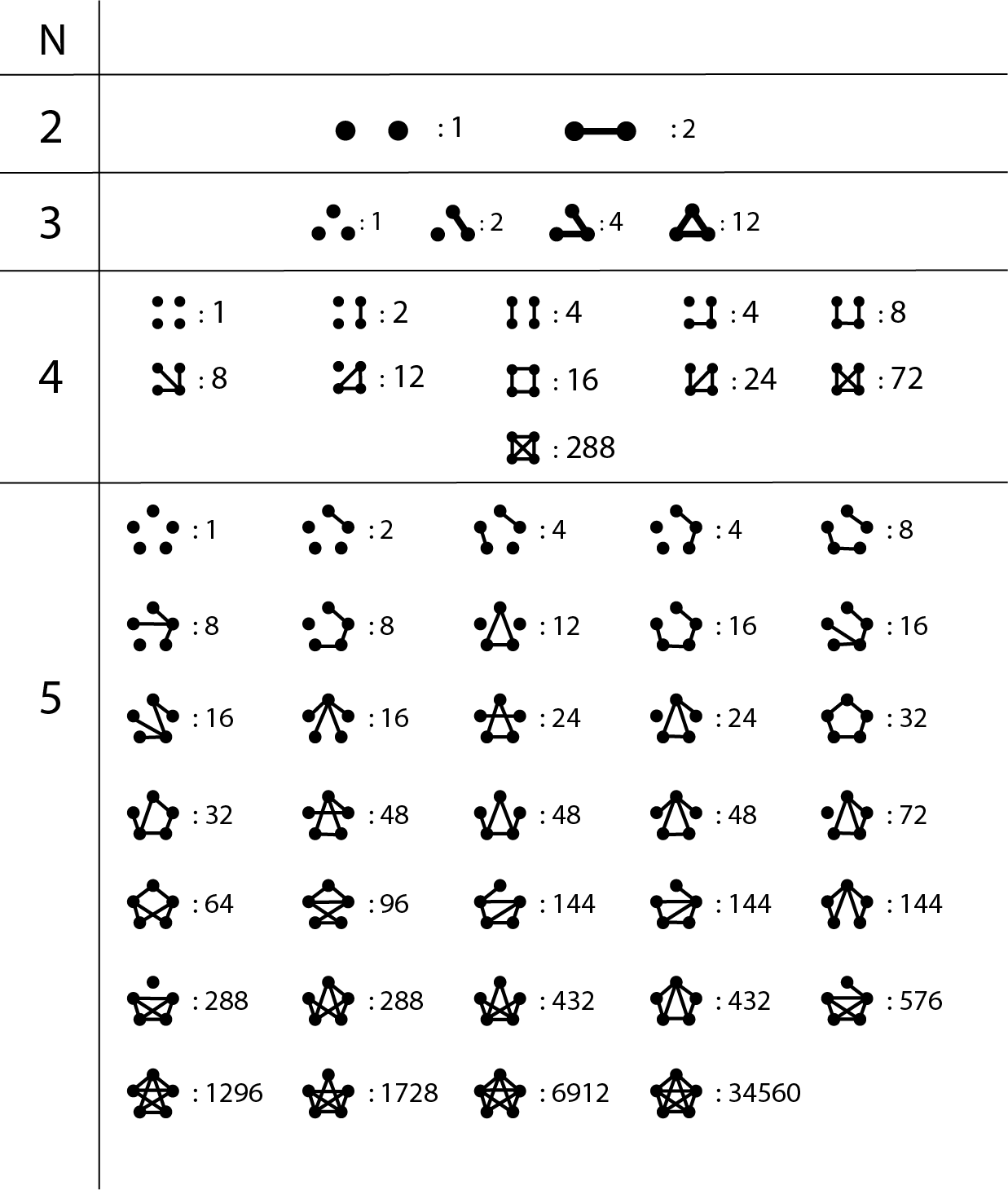} }
\caption{Topologically inequivalent graphs and corresponding coefficients $C_N(\mathbf{m})$ for $\ell=1$ and $N=2,\ldots,5$.}
\label{Fig3}
\end{figure}

For $\ell=1$, the components $m_{i,j}$ of $\mathbf{m}=(m_{i,j})_{i<j}$ are either $0$ or $1$, and one can represent $\mathbf{m}$ by the graph obtained by drawing $N$ nodes $i=1,\ldots,N$ and connecting those pairs of nodes $(i,j)$ where $m_{i,j}=1$. We found that the value of $C_N(\mathbf{m})$ does not depend on the labeling of the nodes and, for this reason, only topologically inequivalent such graphs need to be considered; see Fig.~\ref{Fig2} for these graphs for $N=2,\ldots,5$ and the corresponding coefficients $C_N(\mathbf{m})$. More specifically, our results in Fig.~\ref{Fig2} are all consistent with the following {\bf Conjecture}: {\em Let $f_n$ be the rational numbers determined by the following conditions, 
\begin{equation} 
\prod_{n=2}^N f_n^{\binom{N}{n}} = \prod_{n=2}^N n!\quad (N=2,3,4,\ldots), 
\end{equation} 
i.e., $f_2=2$, $f_3=3/2$, $f_4=8/9$, $f_5=135/128$ etc. Then 
\begin{equation} 
C_N(\mathbf{m}) = \prod_{n=2}^N f_n^{q_n(\mathbf{m})}
\end{equation}  
with $q_n(\mathbf{m})$ the number of cliques of size $n$ in the graph associated with $\mathbf{m}$} (a clique is a fully connected subgraph). 
It is not manifest but true that all $C_N(\mathbf{m})$ obtained in this way are integers $\geq 1$. 
The conjecture implies, in particular, that for the vector $\mathbf{m}_{\rm{max}}$ where all $m_{i,j}=1$ (corresponding to the maximally connected graph where all pairs of nodes are connected by edges), $C_N(\mathbf{m}_{\rm{max}})=\prod_{n=2}^N n!$; we have a proof of the latter result for arbitrary $N\in\Z_{\geq 2}$ (we plan to present this proof elsewhere). 

For general $\ell\in\Z_{\geq 1}$, one can identify $\mathbf{m}$ with multigraphs consisting of $N$ nodes with $m_{i,j}\in\{ 0,\ldots,\ell\}$ counting the number of edges between node $i$ and $j$, and $C_N(\mathbf{m})$ can be interpreted as amplitude corresponding to this multigraph.

\medskip 

\noindent{\bf Appendix C: Further remarks.}  We give various remarks on mathematical aspects of our results. 

\medskip 

\noindent {\bf (i)} We first obtained Proposition~1 by developing a quantum version of the projection method (which is a well-known method used to compute the exact time evolution of the classical variants of CMS systems \cite{kazhdan1978}); this derivation of Proposition~1 is restricted to integer $\ell$ (we plan to present this derivation elsewhere). 
Subsequently, we discovered the elementary proof presented in Appendix~A which applies even to non-integer $\ell$. 

\medskip 

\noindent {\bf (ii)}  It is interesting to note that \eqref{cFN} implies 
\begin{equation} 
\cF_N(x;p) = \sum_{\mathbf{m}} \frac{C_N(\mathbf{m})}{\prod_{i<j} X_{i,j}^{m_{i,j}}}
\end{equation} 
where $\mathbf{m}=(m_{i,j})_{i<j}\in\{0,\ldots,\ell\}^{N(N-1)/2}$. 
We conjecture that all coefficients $C_N(\mathbf{m})$ in this formula are non-negative integers (since this is true in all cases we understand), and that they have interesting combinatorial interpretations (we discuss such an interpretation for $\ell=1$ in Appendix~B(ii)).

\medskip

\noindent {\bf (iii)} It is interesting to note that the eigenfunctions $\Psi_N(x;p)$ \eqref{PsiN} of the Calogero Hamiltonian $H|_{\omega=0}$ are invariant under the exchange $x\leftrightarrow p$ and, for this reason, they are also eigenfunctions of the differential operator 
\begin{equation} 
- \sum_{i=1}^{N}  \frac12 \frac{\partial^2}{\partial p_i^2} + \sum_{i<j} \frac{\ell(\ell+1)}{(p_i-p_j)^2}
\end{equation} 
with eigenvalue $x^2/2$. This property is known as bispectrality in the mathematics literature.  

\medskip 

\noindent {\bf (iv)}  The function $\cF_N(x;p)\ee^{\ii p\cdot x}$ is an eigenfunction of the Calogero Hamiltonian $H|_{\omega=0}$ without harmonic potential, and the corresponding eigenvalue is $p^2/2$. 
It is important to note that this eigenfunction is unphysical  since it is singular; the symmetrization (for odd $\ell$) or anti-symmetrization (for even $\ell$) in \eqref{PsiN} is needed to obtain the non-singular eigenfunctions $\Psi_N(x;p)$ of interest in physics. 
To be more specific: While while $\cF_N(x;p)\ee^{\ii p\cdot x}$ diverges in the limit $x_i\to x_j$  ($i<j$), the (anti-)symmetrized wave function $\Psi_N(x;p)$ vanishes like $\cV_N(x)^{\ell+1}$ in any such limit (this is a consequence of Property (iii) in Proposition~1); this property is not manifest in our explicit formulas for $\Psi_N(x;p)$. 
For $N=2$, this property is implied by well-known representations of the spherical Bessel functions making manifest that $zj_\ell(z)$ vanishes like $z^{\ell+1}$ as $z\to 0$. 
For $N\geq 3$, this property can be proved using an explicit formula for the wave function $\Psi_N(x;p)$  obtained by specializing a known kernel function representation of the eigenfunctions of the hyperbolic Calogero-Sutherland model \cite{hallnas2015} (we plan to present details on this elsewhere). 
We note in passing that this kernel function representation of $\Psi_N(x;p)$ was useful for us during our work.

\medskip 

\noindent {\bf (v)}  The functions $\Psi_N(x;p)$ is equal to Opdam's Bessel functions for the $A_{N-1}$ case \cite{opdam1993}. 
Moreover, $\cF_N(x;p)\ee^{\ii p\cdot x}$ is known as Baker--Akhiezer function in the mathematics literature \cite{chalykh1990}, and explicit formulas for this function were obtained in \cite{chalykh1999,felder2009}. 
We believe our result in Proposition~2 is somewhat simpler from a computational point of view than previously known results.

\medskip 

\noindent {\bf (vi)}  An explicit formula for asymptotically free eigenfunctions of the trigonometric Macdonald-Rujsenaars operators for $N=3$ that is similar Proposition~2 was obtained by Noumi and Shiraishi \cite[Theorem 7.2]{noumi2012}; it would be interesting to understand the precise relation between their and our result. 
In particular, we hope that a future extension of  Proposition~2 to $N>3$ could help to find an extension of the Noumi-Shiraishi result to $N>3$.

\medskip 

\noindent {\bf (vii)} It would be interesting to extend our results to non-integer $\ell\geq -1/2$ where the Calogero model is known to describe anyons \cite{polychronakos2006}.
We expect that, for non-integer $\ell$, the series corresponding to \eqref{cFN} become infinite and, for this reason, nontrivial issues of convergence arise. 
It would also be interesting to find other CMS-type models were the eigenfunctions can be expressed as finite sums of products of two-body functions.


\begin{thebibliography}{10}
\bibitem{remark2}
We analytically continue time, $t$, so that $\re(\ii t)>0$.
\bibitem{remark3}
One can also verify directly that $\ee^{\ii p\cdot x}\cF_2(x;p)$ is an
  eigenfunction of $H|_{N=2,\omega=0}$.
\end{thebibliography}
\end{document}